\documentclass[doc,floatsintext,natbib]{apa7}
\usepackage[english]{babel}
\usepackage[utf8]{inputenc}
\usepackage{amsmath}
\usepackage{graphicx}
\usepackage{subcaption}
\usepackage{doi}
\usepackage{booktabs}
\usepackage{placeins}
\usepackage{enumitem}
\usepackage{xcolor}
\usepackage{setspace}
\usepackage{hyperref}

\title{Subgroup detection in linear growth curve models with generalized linear mixed model (GLMM) trees}
\shorttitle{Recursive partitioning of linear growth curve models}
\authorsnames{Marjolein Fokkema$^1$, Achim Zeileis$^2$}
\authorsaffiliations{{$^1$Leiden University}, {$^2$Universit\"at Innsbruck}}

\abstract{Growth curve models are popular tools for studying the development of a response variable within subjects over time. Heterogeneity between subjects is common in such models, and researchers are typically interested in explaining or predicting this heterogeneity. We show how generalized linear mixed effects model (GLMM) trees can be used to identify subgroups with differently shaped trajectories in linear growth curve models. Originally developed for clustered cross-sectional data, GLMM trees are extended here to longitudinal data. The resulting extended GLMM trees are directly applicable to growth curve models as an important special case. In simulated and real-world data, we assess the performance of the extensions and compare against other partitioning methods for growth curve models. Extended GLMM trees perform more accurately than the original algorithm and LongCART, and similarly accurate as structural equation model (SEM) trees. In addition, GLMM trees allow for modeling both discrete and continuous time series, are less sensitive to (mis-)specification of the random-effects structure and are much faster to compute.\\}

\usepackage{Sweave}
\begin{document}
\Sconcordance{concordance:_Partitioning_GCMs_with_GLMM_trees.tex:_Partitioning_GCMs_with_GLMM_trees.Rnw:1 %
32 1 1 0 9 1 1 5 1 16 8 1 1 11 1 2 17 1 1 28 1 2 6 1 1 11 9 0 1 2 5 1 1 %
16 67 1 1 8 1 2 14 1 1 6 1 2 53 1 1 70 3 1 1 7 1 2 57 1 1 36 1 35 1 23 %
1 9 6 1 1 88 1 2 19 1 1 68 28 0 1 2 40 1 1 10 1 1 1 29 2 1 1 33 3 1 1 %
64 1 2 9 1 1 23 5 1 1 8 12 1 1 18 26 0 1 2 6 1 1 117 9 1 1 22 1 2 82 1 %
1 88 1 2 1 1 1 33 1 2 1 1 1 29 1 2 6 1 1 25 3 1 1 39 23 0 1 2 1 14 4 1 %
1 57 1 2 1 1 1 25 1 2 1 1 1 27 1 2 46 1 1 17 1 2 1 1 1 15 1 2 1 1 1 15 %
1 2 1 1 1 15 1 3 1 1 1 15 1 2 12 1 1 16 1 2 1 1 1 16 1 2 1 1 1 15 1 2 1 %
1 1 15 1 2 1 1 1 18 1 2 21 1 1 184 5 1 1 43 151 1 1 28 2 1 1 21 1 1 1 %
64 10 1}

\maketitle
\setlength{\tabcolsep}{3pt}

\section{Introduction}
\label{sec:Introduction}

Development over time is of prime interest in psychological research. For example, in educational studies researchers may want to model student's academic development over time; in clinical studies researchers may want to model patients' symptoms over time. Mixed-effects or latent-variable models can be used to model such trajectories and allow for explaining heterogeneity with covariates of a-priori known relevance \citep[e.g., ][]{NeisyMatt18}. However, when these covariates are not known in advance, methods for identifying them are needed. 

\begin{figure}[b]%
\caption{Growth curves of science ability.}
\begin{subfigure}{0.65\textwidth}%
\includegraphics{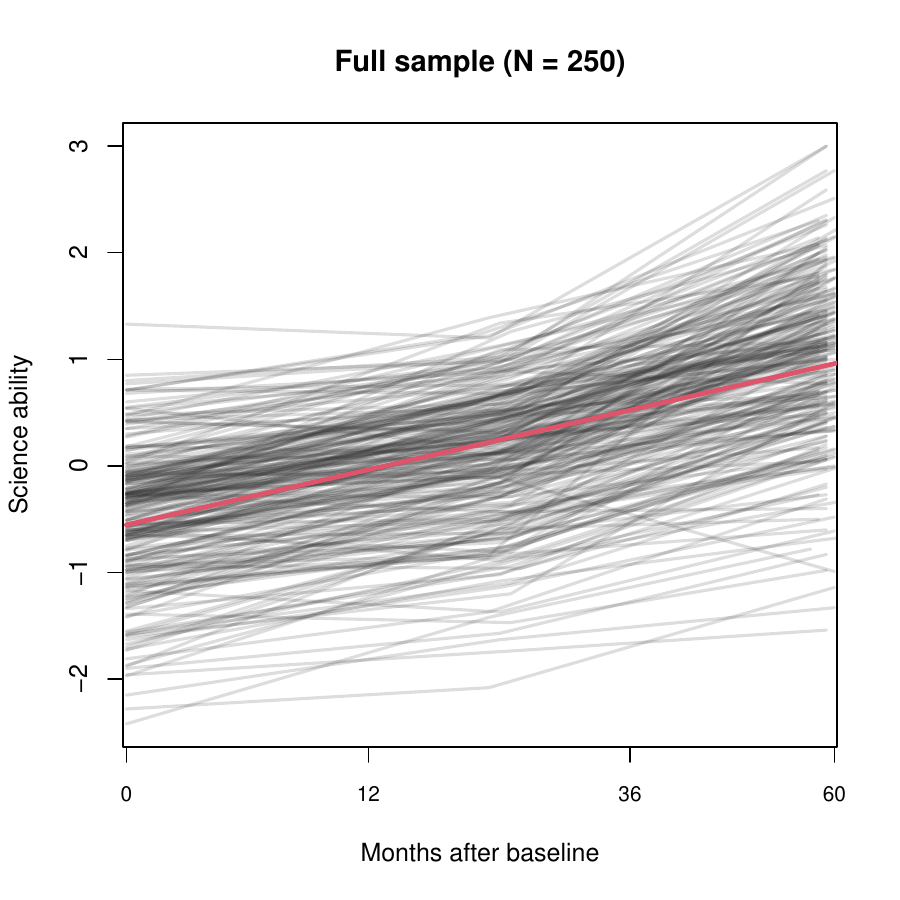}
\\\footnotesize{\textit{Note.} Gray lines depict observed individual trajectories, red line depicts average growth curve as estimated with a linear mixed-effect model, comprising a fixed effect of time and a random intercept with respect to individuals. The $x$-axis is not linear in the number of months, because time was scaled as $(\text{\# of months})^\frac{2}{3}$ in order to obtain approximately linear trajectories.}
\end{subfigure}%
\label{fig:global_trajectories}
\end{figure}

As an example, trajectories of science knowledge and skills from a sample of 250 children are depicted in Figure~\ref{fig:global_trajectories}. The children were assessed at three timepoints across grades 3 through 8\footnote{Further details on the source of these data are provided in Study~III.}. The red line depicts the estimated average trajectory, while the gray lines depict individual trajectories. The gray lines reveal substantial variability between the children, both in initial levels and growth over time. An obvious research aim would be to identify covariates that can explain or predict this heterogeneity. 

\subsection{Recursive Partitioning Methods for Growth Curve Models}

\begin{figure}%
\caption{Partitioned growth curves of science ability.}
\begin{subfigure}{0.9\textwidth}%
\includegraphics{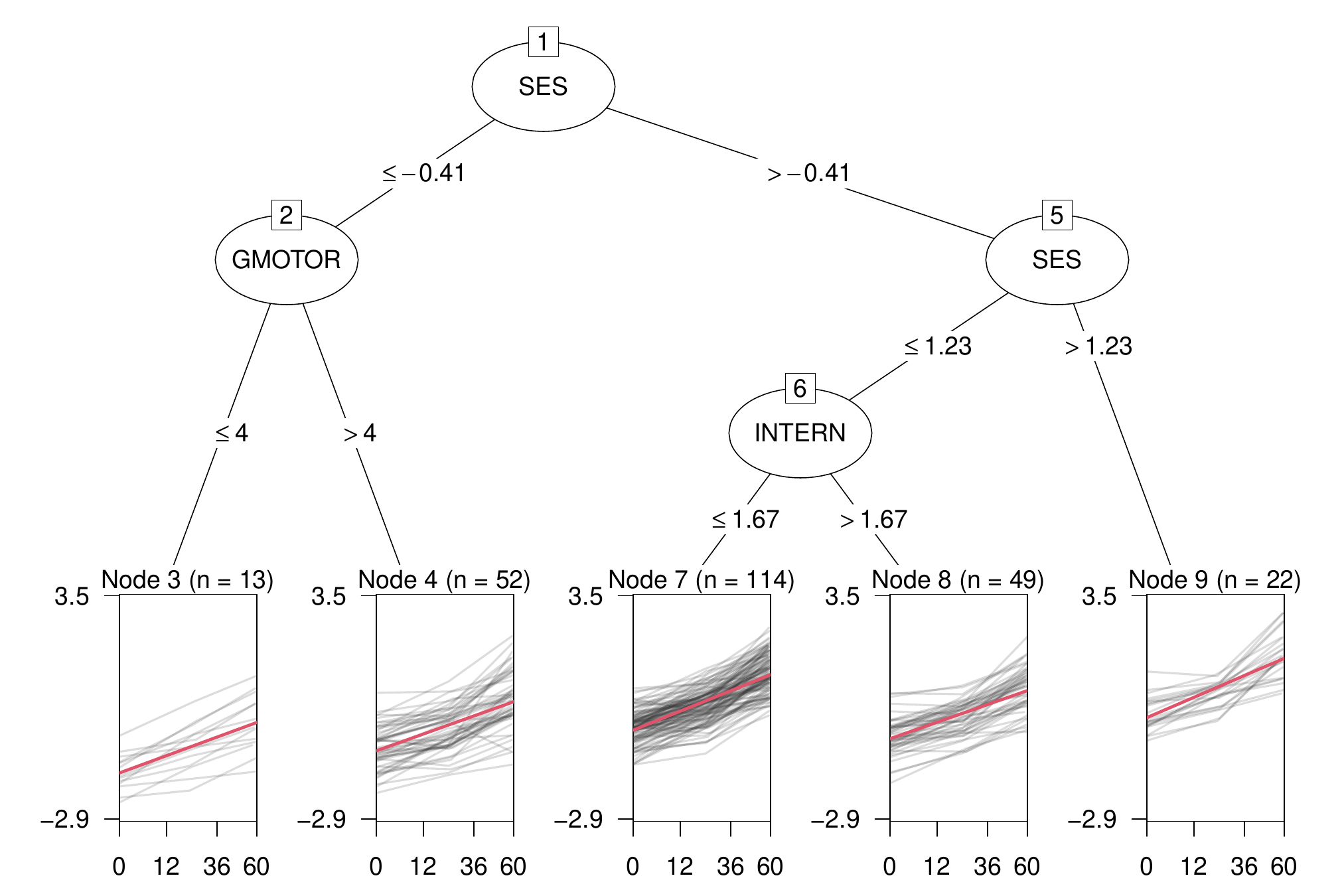}
\label{subfig:tree}
\end{subfigure}%
~
\begin{subfigure}{0.25\textwidth}%
\vspace{-2.95cm}
\hspace{-2.5cm}
\scriptsize
\begin{tabular}{ccc}
\toprule
Node & Intercept & Slope\\
\midrule
3 & $-1.576$ & $0.094$\\
4 & $-0.942$ & $0.092$\\
7 & $-0.350$ & $0.104$\\
8 & $-0.595$ & $0.090$\\
9 & $\:\:$ $0.009$ & $0.111$\\
\bottomrule
\end{tabular}\label{subfig:coefs}
\end{subfigure}
\\\footnotesize{\textit{Note.} The $x$-axes represent the number of months after the baseline assessment, $y$-axes represent science ability. Gray lines depict observed individual trajectories, red lines depict average growth curve within each terminal node, as estimated with a linear mixed-effect model comprising node-specific fixed effects of time and a random intercept with respect to individuals. The $x$-axis is not linear in the number of months, because time was scaled as $(\text{\# of months})^\frac{2}{3}$ in order to obtain approximately linear trajectories. The table presents numerical estimates of fixed intercepts and slopes.}
\label{fig:LMM_tree_r}
\end{figure}

Recursive partitioning methods (RPMs), also known as ``trees``, allow for identifying relevant predictors from a potentially (very) large number of covariates. Figure~\ref{fig:LMM_tree_r} shows an example tree, which identified socio-economic status (SES), gross motor skills (GMOTOR) and internalizing problems (INTERN) from a set of 11 socio-demographic and behavioral characteristics of the children, assessed at baseline. Five subgroups were identified, corresponding to the terminal nodes of the tree, each with a different estimate of the fixed intercept and slope. Groups of children with higher SES also have higher intercepts, indicating higher average science ability. The group of children with lower SES (node~2) is further split based on gross motor skills, with higher motor skills resulting in a higher intercept. The group of children with intermediate levels of SES (node~6) is further split based on internalizing problems, with lower internalizing problems resulting in a higher intercept. The two groups (or nodes) with higher intercepts also have higher slopes, indicating that children with higher ability also gain more ability over time.

The tree in Figure~\ref{fig:LMM_tree_r} has been estimated with generalized linear mixed-effects model trees \citep[GLMM trees; ][]{FokkySmit18}. GLMM trees were originally proposed for subgroup detection in clustered cross-sectional studies, where subjects are nested in treatment centers, classrooms and/or geographical areas, for example. In the current paper we extend GLMM trees, so that they can be applied to partitioning LGCMs. The general idea of GLMM trees is appealing for subgroup detection in almost any type of mixed-effects model. Compared to clustered cross-sectional data, however, longitudinal data may require a different estimation approach: The variance of random effects tends to be higher with longitudinal data, and the predictors of interest tend to be measured at higher levels (e.g., time-invariant covariates). In this paper, we propose and test two extensions of GLMM trees that account for these characteristics. We focus on the specific use case of partitioning LGCMs, but the extensions are critical for a wider range of settings where covariates are measured at higher levels and/or where the random effects have substantial variance.

There are several alternative RPMs that can be used to partition linear growth-curve models (LGCMs): GUIDE \citep{Loh02}, longRPart \citep{AbdoyLeBl02}, GEE-based decision trees \citep{Lee05}, longitudinal interaction trees \cite[IT; ][]{SuyMene11}, SEM trees \citep{BranyOert13}, mixed-effects longitudinal trees \citep[MELT; ][]{EoyCho14} and LongCART \cite{KundyHare19}. Further, the longRPart2 \citep{StegyJaco18} and IT-LT \citep{WeiyLiu20} methods allow for subgroup detection in non-linear growth curve models\footnote{Both IT methods specifically target subgroups with different time-by-treatment interactions, so are not generally applicable for partitioning growth curve models.}.

The main characteristic that sets GLMM trees apart from other methods for partitioning LGCMs is its local-global estimation approach: GLMM trees do not fit a full parametric model in each of the subgroups defined by the terminal nodes of the tree. Instead, fixed-effects parameters are estimated \textit{locally}, using the observations within a terminal node, and the random-effects parameters are estimated \textit{globally}, using all observations. This local-global estimation approach was first proposed by \cite{HajjyBell11} and \cite{SelaySimo12} for trees with constant fits (i.e., intercepts only) in the terminal nodes. With GLMM trees, the approach was generalized to allow for GLMs with any number of parameters in the terminal nodes, thus allowing for non-Gaussian responses and targeted detection of a wide range of possible interaction effects in mixed-effects models \citep{FokkySmit18}.

Other methods for partitioning LGCMs take a fully local estimation approach: Within every node or subgroup defined by the terminal nodes, a full parametric model is estimated based on the observations in that subgroup only. This fully local estimation approach provides more flexibility, but yields higher computational burden and model complexity. In contrast, GLMM trees estimate a (much) lower number of random-effects parameters, which likely reduces overfitting and improves stability and generalizability of the results. Furthermore, the fully local estimation requires possible partitioning variables to be measured at the highest level of nesting, while GLMM trees' local-global estimation approach allows partitioning variables to be measured at any level.

The computational advantage of GLMM trees is strongest compared to longRPart, longRPart2, IT-LT and LRT-based SEM trees. These methods employ an exhaustive split detection procedure, where for every possible split point in the current node, the full parametric model needs to be re-estimated in the two resulting nodes. To choose the optimal split, the splitting criterion (such as a $p$-value from a likelihood-ratio test) is derived from these two models. Not only does this bring a heavy computational load, it also introduces a selection bias towards covariates with a larger number of possible cutpoints \citep{ShihyTsai04, Shih04}. LongCART, MELT, GEE-based decision trees and score-based SEM trees also fit full parametric models in each of the nodes, but do not require model refitting for cutpoint selection; they employ the predictions or residuals from the fitted model in the current node for selecting the best split. This reduces computational load, while it also allows for separating variable and cutpoint selection, thus preventing selection bias. The GLMM tree algorithm shares these advantages, because it also employs a two-step approach to split selection.

Given their unbiased variable selection, lower model complexity and computational burden, GLMM trees might be particularly useful for subgroup detection in LGCMs. The next section explains how GLMM trees are estimated and propose adjustments for partitioning longitudinal trajectories. Next, the performance of the proposed adjustments is evaluated: In Study~I, we assess performance in simulated datasets, in Study~II, we compare performance of GLMM trees with that of two other methods: SEM trees and LongCART. In Study~III, we assess performance of the proposed adjustments in existing datasets on children's development of reading, math and science abilities. In the Discussion, we summarize our findings and discuss implications.




\FloatBarrier
\section{Estimation of GLMM trees and Adaptations for Longitudinal Data}
\label{sec:estimation}

In the GLMM tree model \citep{FokkySmit18}, expectation $\mu_i$ of outcome vector $y_i$ is modeled through a linear predictor and suitable link function:
\begin{eqnarray}
\label{eq:expected_value}
E[y_i | X_i] & = & \mu_i, \\
\label{eq:GLMMtree}
g(\mu_{i}) & = & X_{i} \beta_{j} + Z_{i} b_{i}
\end{eqnarray}
Throughout this paper we focus on the case with a continuous, normally-distributed response $y_i$ with constant variance $\sigma_\epsilon$. Therefore, the identity function can be used for the link $g$ but generalizations to other response variable types within the GLM are straightforward. In the general notation above, $X_i$ is the $n_i \times (p+1)$ fixed-effects design matrix for subject~$i$ ($i = {1, \dots, N}$), comprising $p$ regressors plus one column of 1s for the intercept. In the following, we assume that time is the predictor variable of interest (i.e., $p = 1$), where the number $n_i$ and spacing of observed timepoints may differ between subjects. The fixed-effects parameters $\beta_j$ (here, intercept and time slope) in GLMM trees are node-specific, i.e., their value depends on the subgroup/node~$j$ into which subject~$i$ falls. As in a traditional GLMM, $Z_i$ is the random-effects design matrix for subject $i$, comprising a subset of columns of $X_i$, and $b_i$ is the corresponding vector of random effects for subject $i$. Finally, $b_i$ is assumed to follow a (possibly multivariate) normal distribution with mean zero and (co)variance $\Sigma$.

The parameters of a traditional GLMM can be estimated, among other techniques, by maximum likelihood (ML) or restricted ML (REML). Thus, when it is known into which node~$j$ each subject~$i$ falls, the GLMM specified by Equation~\ref{eq:GLMMtree} can be fitted ``as usual'', yielding \emph{local} subgroup-specific fixed-effect estimates~$\hat \beta_j$ and \emph{global} random-effect estimates~$\hat b_i$. To infer the subgroup membership for all observations~$i$, the random-effect estimate is treated as a known offset and a GLM tree is estimated using the model-based (MOB) recursive partitioning algorithm of \cite{ZeilyHoth08}. The overall GLMM tree model is then estimated by alternating between estimating the partition (i.e., subgroups or terminal nodes $j$), and estimating the random- and fixed-effects parameters, as per the following algorithm:
\begin{enumerate}
\setlength\itemsep{0em}
	\setcounter{enumi}{-1}
	\item Initialize by setting step $r = 0$ and all random-effect estimates $\hat{b}_{i,(r)} = 0$.
	\item Set $r = r+1$. Fit a GLM tree using $Z_{i} \hat{b}_{i,(r-1)}$ as an offset, yielding the partition $j_{(r)}$.
	\item Fit the mixed-effects model $g(\mu_{i}) = X_{i} \beta_{j, (r)} + Z_{i} b_{i, (r)}$ with the partition $j_{(r)}$ from Step~1. Extract the random-effect estimates~$\hat{b}_{i,(r)}$ from the fitted model.
	\item Repeat Steps~1 and~2 until convergence.
\end{enumerate}
This initialization simply assumes zero random effects. Convergence of the algorithm is monitored through the log-likelihood of the mixed-effects model fitted in Step~3. Typically, this converges when the partition~$j_{(r)}$ from the GLM tree is the same as $j_{(r-1)}$ from the previous step.

The following two subsections describe alternative approaches for the initialization in Step~0 and for fitting the GLM tree in Step~1. Each subsection first reviews the well-established methods and then proceeds to discuss modifications that may improve performance when partitioning longitudinal data.

\subsection{Initialization}

Previous publications on mixed-effects recursive partitioning find that initializing the random-effect estimates with zero yields accurate estimates of subgroup memberships and the final models \citep{HajjyBell11, HajjyBell14, HajjyLaro17, SelaySimo12, FuySimo15, FokkySmit18}. \cite{SelaySimo12} assessed the impact of different initialization values and found only minor differences that decreased with increasing sample size. In \cite{FokkySmit18}, we found initializing estimation of GLMM trees with zero random effects performed well in cross-sectional clustered data. With longitudinal data, however, random effects tend to be more pronounced: Repeated measures on the same subjects tend to be correlated more strongly than observations nested within the same unit in cross-sectional data. If random effects are sizable, the initial assumption of zero random effects could provide an unrealistic starting point that may be difficult to overcome in subsequent iterations. Our expectation is that for partitioning LGCMs, initializing estimation with the random effects instead of the subgroup structure may improve subgroup recovery. Specifically, that means the algorithm starts by estimating the classic version of the mixed-effects model from Equation~\ref{eq:GLMMtree} with just one set of fixed-effects coefficients $\beta$ and all subjects in a single group.

The alternative initialization step is thus:
\begin{enumerate}
\setlength\itemsep{0em}
\renewcommand{\labelenumi}{\arabic{enumi}$'$.}
	\setcounter{enumi}{-1}
	\item Initialize by setting step $r = 0$ and fit the mixed-effects model $g(\mu_{i}) = X_{i} \beta + Z_{i} b_{i, (r)}$ to the full sample. Extract the random-effect estimates~$\hat{b}_{i,(r)}$ from the fitted model.
\end{enumerate}

\begin{figure}%
\caption{GLMM tree estimated by initializing with zero random effects.}
\begin{subfigure}{1.2\textwidth}
\includegraphics{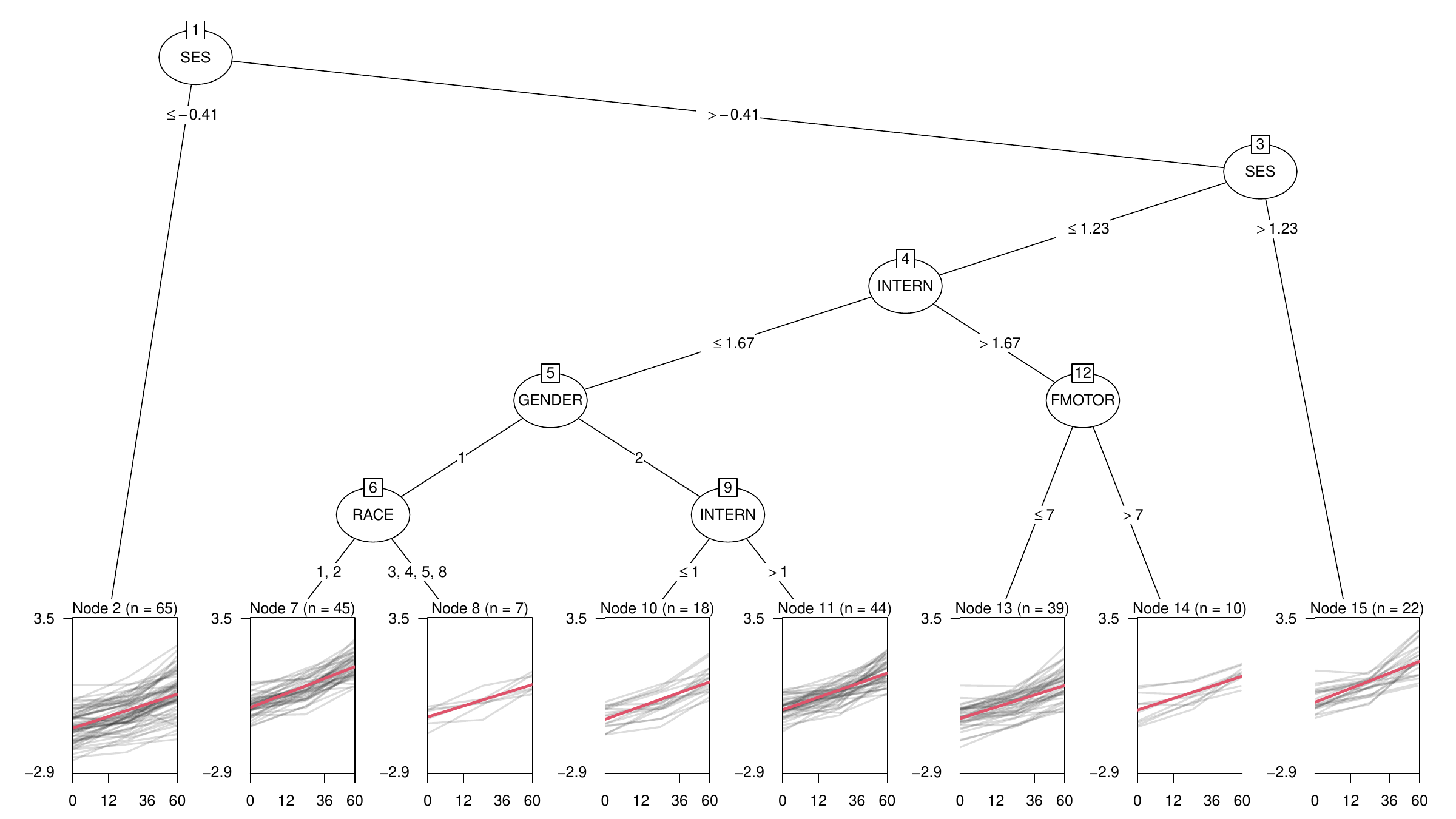}
\end{subfigure}
\\\footnotesize{\textit{Note.} The $x$-axes represent the number of months after the baseline assessment, the $y$-axes represent science ability. Gray lines depict observed individual trajectories. Red lines depict the average growth curve within each terminal node, as estimated with a linear mixed-effect model with a node-specific fixed effect of time and random intercepts estimated with respect to individuals.}
\label{fig:LMM_tree}
\end{figure}%

To illustrate, we applied both initialization approaches to the dataset from Figure~\ref{fig:global_trajectories}. In fact, the tree presented in Figure~\ref{fig:LMM_tree_r} was estimated by initializing with the random effects. Initializing estimation assuming zero random effects resulted in the tree in Figure~\ref{fig:LMM_tree}. The split based on gross motor skills in Figure~\ref{fig:LMM_tree_r} was not implemented in  Figure~\ref{fig:LMM_tree}, while additional splits were implemented based on gender, race, internalizing problems and fine motor skills. Considering the relatively large number of subgroups in  Figure~\ref{fig:LMM_tree_r}, some with relatively small sample sizes, this tree may overfit and not generalize well to other samples.

\FloatBarrier
\subsection{Partitioning}

\begin{figure}%
\caption{GLMM tree estimated with cluster-level parameter stability tests.}
\includegraphics{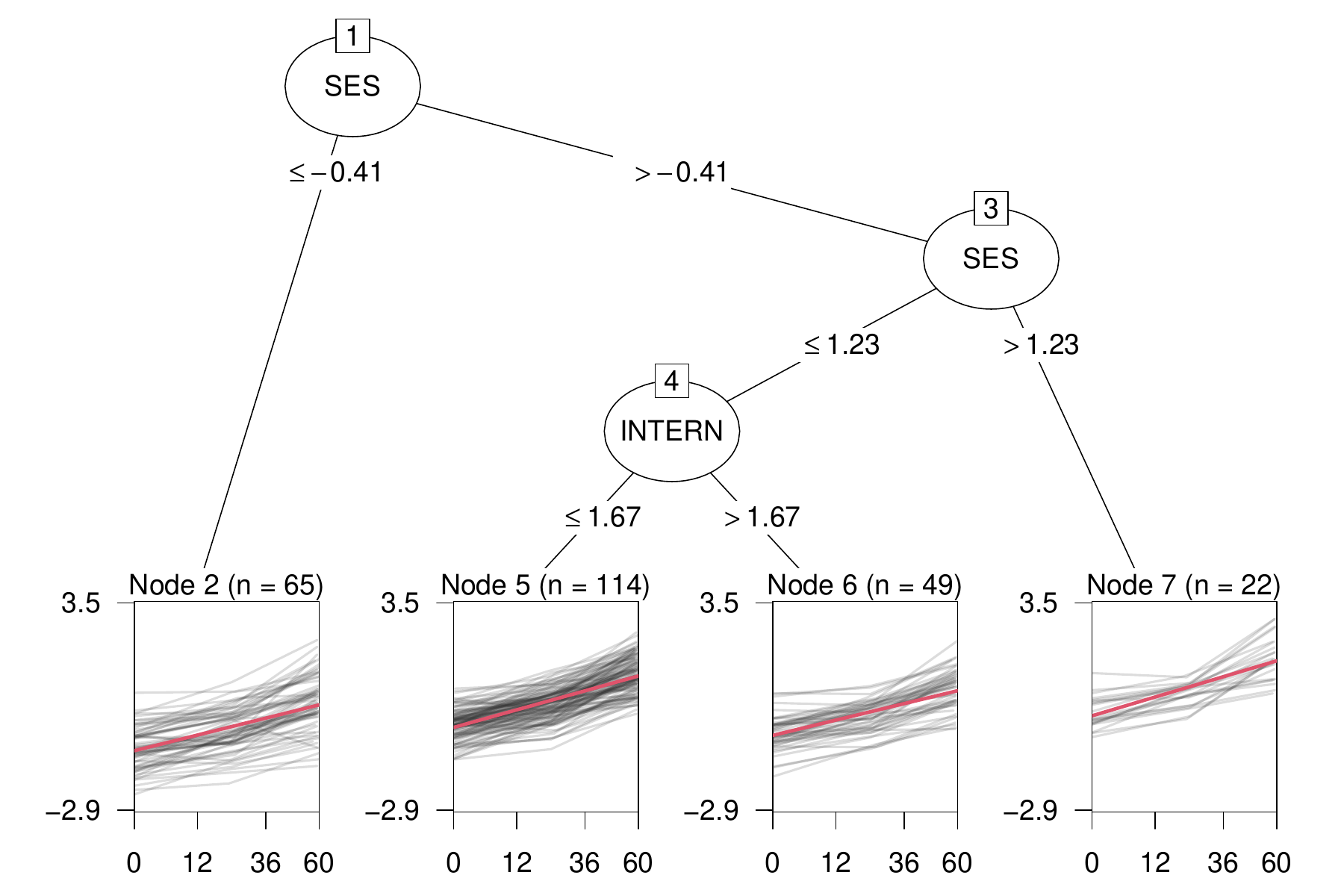}
\\\footnotesize{\textit{Note.} The $x$-axes represent the number of months after the baseline assessment, $y$-axes represent science ability. Gray lines depict observed individual trajectories. Red lines depict the average growth curve within each terminal node, as estimated with a linear mixed-effect model with a node-specific fixed effect of time and random intercepts estimated with respect to individuals.}
\label{fig:LMM_tree_c}
\end{figure}%

The subgroup structure in Step~1 is estimated by a GLM tree using the general model-based recursive partitioning (MOB) algorithm of \cite{ZeilyHoth08}. Here, we give a general overview and then comment on the aspects of the algorithm that may be particularly relevant for LGCMs. In the case of GLMs (with random effects held constant in an offset), the MOB algorithm cycles iteratively through the following steps: 
\begin{enumerate}
\setlength\itemsep{0em}
\renewcommand{\labelenumi}{(\alph{enumi})}
	\item Fit the GLM to all observations in the current subgroup. 
	\item Test for instability of the GLM parameters with respect to each of the partitioning variables.
	\item If there is some overall parameter instability, split the subgroup with respect to the partitioning variable associated with the highest instability.
	\item Repeat Steps (a) through (c) in each of the resulting subgroups.
\end{enumerate}
Parameter stability in Step~(b) is tested using the the \textit{scores} (gradient contributions) from the GLM fitted in Step~(a). Under correct specification of the model and mild regularity conditions, the scores have an expected value of 0. The parameter stability tests evaluate whether the scores fluctuate randomly around this mean of 0, or exhibit systematic deviations when ordered by the values of a covariate available for partitioning. For continuous covariates $u_k$ (or ordered covariates with a large enough number of unique values), this involves computing the following cumulative score process $W_k(t)$ with respect to each potential partitioning variable \citep{ZeilyHoth08}:
\begin{eqnarray}
\label{eq:efp}
W_{k}(t) = \hat{J}^{-1/2} n_{j}^{-1/2} \sum^{[n_jt]}_{i=1}{\hat{\psi}}_{\sigma(u_{ik})}
\end{eqnarray}
where $\hat{J}$ is a suitable estimate of the covariance matrix of the parameter estimates, and $n_j$ gives the number of observations in the current subgroup. Further, $\hat{\psi}_{\sigma(u_{ik})}$ denotes the scores evaluated at the parameter estimates, with subscript {$\sigma(u_{ik})$} denoting their ordering by the values of partitioning covariate $u_k$. Note that $0 \leq t \leq 1$, thus $n_jt = 1$ for an observation associated with a unique minimum on the partitioning variable, and $n_jt = n_j$ for an observation with a unique maximum. 

From the cumulative score process $W_k(t)$, a range of test statistics can be derived which capture increased fluctuations (beyond the random fluctuation under parameter stability). For numerical partitioning variables, a maximum Lagrange multiplier test statistic can be computed, which takes the maximum of the squared Euclidean norm of $W_k(t)$, weighted by its variance \citep{ZeilyHorn07}. This statistic is referred to as the \textit{supLM} statistic, and is asymptotically equivalent to the maximum of likelihood-ratio statistics. Approximate asymptotic $p$-values for the \textit{supLM} statistic can be computed with the method of \cite{Hans97}. Categorical covariates do not provide an implicit ordering and scores are therefore binned at each level of the covariate. From these, a test statistic is computed that does not depend on the ordering of the levels \citep{MerkyFan14}. 

When partitioning longitudinal data, covariates will often be measured at the subject level (i.e., time-invariant covariates), which should be accounted for in computing the estimated covariance matrix $\hat{J}$. In general, this computation makes use of the scores. By summing the scores within clusters prior to computation of the covariances, so-called \textit{clustered} covariances are obtained, which account for dependence between observations within the same cluster \citep{ZeilyKoll20}. This resembles a GEE-type approach with an independence correlation structure. Our expectation is that in partitioning LGCMs, use of clustered covariances in the parameter stability tests will improve subgroup recovery.

The tree in Figure~\ref{fig:LMM_tree_c} was estimated using cluster-level parameter stability tests. Application of GLMM trees with both cluster-level parameter stability tests and random-effects initialization yielded the exact same tree structure and parameter estimates. Compared to Figures~\ref{fig:LMM_tree_r} and \ref{fig:LMM_tree}, the cluster-level parameter stability tests provided the most parsimonious tree structure thus far. The estimated variances of the random intercept were 0.21 (Figure~\ref{fig:LMM_tree_r}, 0.20 (Figure~\ref{fig:LMM_tree}) and 0.23 (Figure~\ref{fig:LMM_tree_c}, indicating that the more parsimonious the tree structure, the more variance will be captured by the random effects.

In the next sections, we assess performance of the original algorithm and proposed adaptations through more extensive empirical evaluations.



\FloatBarrier
\section{Study~I: Assessment of Subgroup Recovery}

\subsection{Method}

\subsubsection{Data generation}

\begin{figure}[tb]
\caption{Design of subgroups and fixed effects.}
\includegraphics{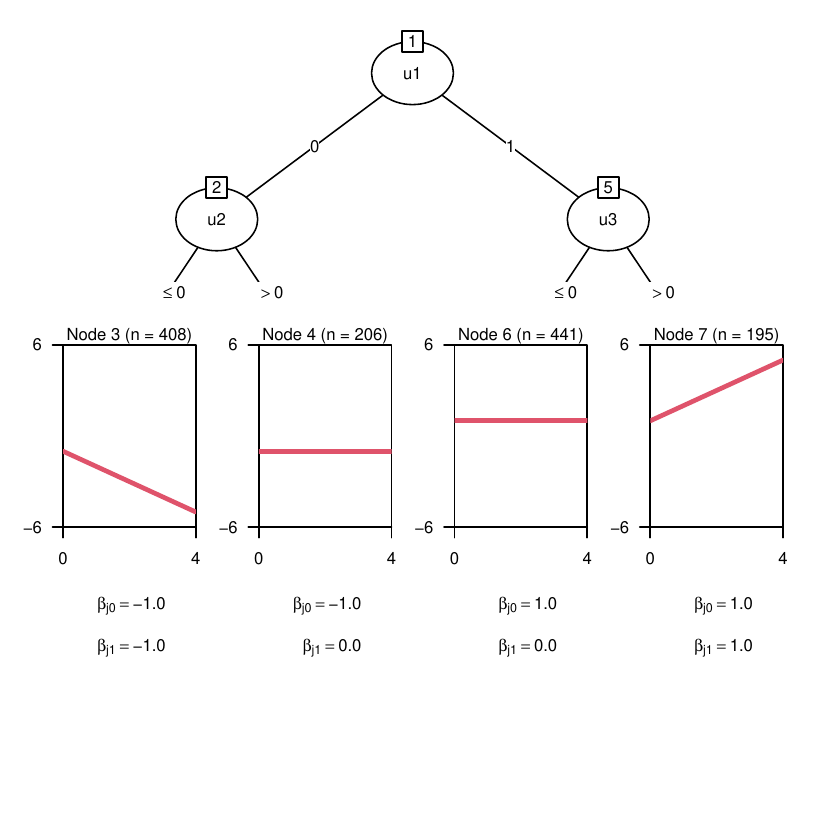}
\label{fig:design_tree}
\end{figure}

We simulated data according to the subgroup structure depicted in Figure~\ref{fig:design_tree}. Every dataset comprised four non-overlapping subgroups, corresponding to the terminal nodes of the tree in Figure~\ref{fig:design_tree}. The subgroups are defined by the three true partitioning variables: $u_1$, $u_2$ and $u_3$. All partitioning variables were generated from a standard normal distribution with $\mu = 0$ and $\sigma^2 = 25$. To allow for assessing possible selection bias toward partitioning variables with a larger number of possible cutpoints, variable $u_1$ was transformed to a binary factor, with values below the mean set to 0 and values above the mean set to 1. The response was computed as:
$$y_{i} =  X_{i} \beta_{j} + Z_{i} b_i + \epsilon_{i},$$
where $\beta_j$ corresponds to the fixed effects in terminal node $j$ of which subject $i$ is part. $\beta_j$ values are reported below the terminal node panels in Figure~\ref{fig:design_tree}. The fixed- and random-effects design matrices $X_i$ and $Z_i$ are identical, each comprising two columns: a vector of 1s for the intercept, and a vector of timepoints. The same set of timepoints was generated for all subjects: ${0, 1, 2, 3, 4}$\footnote{Although GLMM trees are not restricted to have the same set of timepoints for each subject, this design allowed for comparison with SEM trees in Study~II.}. Values of $b_i$ (random intercepts and slopes) were generated from a multivariate normal distribution with mean zero and a $2 \times 2$ diagonal covariance matrix $\Sigma$, the diagonal entries determined by the level of the data-generating design described below. Values of $\epsilon_{i}$ were independently generated from a normal distribution with $\mu = 0$ and $\sigma^2 = 5$.   

We varied the following five data-generating characteristics:
\begin{itemize}
\setlength\itemsep{0em}
\setlength{\itemindent}{0.2in}
\item Number of subjects: small ($N = 100$) or large ($N = 250$).
\item Variance of the random intercept: small ($\sigma_{b_0}^2 = 1$) or large ($\sigma_{b_0}^2 = 4$).
\item Variance of the random slope: small ($\sigma_{b_1}^2 = 0.1$) or large ($\sigma_{b_1}^2 = 0.4$).
\item Number of noise variables: small ($p = 5$) or large ($p = 25$).
\item Intercorrelation between partitioning variables: absent ($\rho = 0$) or present ($\rho = 0.3$).
\end{itemize}
A full factorial design was employed, yielding $2^5 = 32$ cells of the design; 100 repetitions were performed per cell. All data generation and analysis was performed in R \citep[version 4.1.2; ][]{R22}.

\subsubsection{Model fitting}

We applied ten different fitting approaches to every generated dataset. Each variation combines one of three \emph{random-effects specifications} (none vs. intercepts vs. intercepts+slopes) with one of two \emph{random-effect initializations} (if any; all zero vs.\ full sample estimates) and one of two \emph{covariance specifications in the parameter instability tests} (classic vs.\ clustered):
\begin{itemize}
  \item None: $\hat{\sigma}_{b_0} = \hat{\sigma}_{b_1} = 0$. Random effects are not estimated and their variances thus fixed to 0, yielding linear model (LM) trees with fixed effects only and the following variations of covariance specifications:
  \begin{itemize}
      \item Default: Classic observation-level covariances.
      \item Alternative: Clustered covariances. 
  \end{itemize}
  
  \item Intercepts: $\hat{\sigma}_{b_0} > 0; \hat{\sigma}_{b_1} = 0$. This yields linear mixed-effects model (LMM) trees in which only the variance of the random intercept was freely estimated and the variance of the random slope was fixed to 0. The four variations considered are the following:
  \begin{itemize}
    \item Default: Classic observation-level covariances in the parameter stability tests and random-effect initialization with all zeros (original step 0.). 
    \item Alternative: Clustered covariances in the parameter stability tests.
    \item Alternative: Random-effect initialization with the full-sample estimates (alternative step 0$'$.). 
    \item Alternative: Clustered covariances and random-effect initialization with the full-sample estimates.
  \end{itemize}
  
  \item Intercepts and slopes: $\hat{\sigma}_{b_0} > 0; \hat{\sigma}_{b_1} > 0$. This yields LMM trees in which the variance of both the random intercept and slope were freely estimated. The four variations considered are the same as for the intercept-only LMM trees.
\end{itemize}
To fit LM trees, we used package \textbf{partykit} \citep[version 1.2-15; ][]{HothyZeil15}. To fit LMMs, we used package \textbf{lme4} \cite[version 1.1-29; ][]{BateyMach15}. To fit LMM trees we used package \textbf{glmertree} \citep[version 0.2-0; ][]{FokkySmit18}. To compute clustered covariances, we used package \textbf{sandwich} \citep[version 3.0-1; ][]{ZeilyKoll20}. We employed the outer product of gradients method to compute covariances, thus employing only the meat of the sandwich estimator for the covariances.

\subsubsection{Evaluation of performance}

We evaluated tree accuracy by counting the number of splits in each tree, and computing the standard (SD) and mean absolute deviation (MAD) from the true tree size (3 splits). Trees with $> 3$ splits are indicative of Type-I error, while trees with $< 3$ splits are indicative of Type-II errors (i.e., power too low to detect the true partitioning variables). We also assessed whether the variable selected for the first split in every tree was the true first splitting variable. 

\FloatBarrier
\subsection{Results}

\begin{figure}[b]
\caption{Tree size distributions for LM trees (left panel) and LMM trees (middle and right panel).}
\begin{subfigure}{1.25\textwidth}
\includegraphics{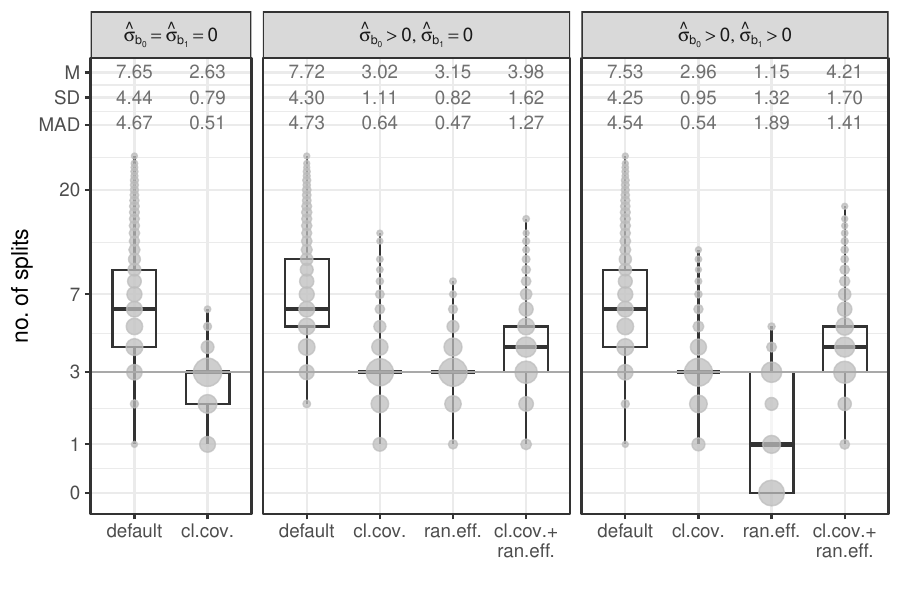}
\end{subfigure}%
\vspace{-0.5cm}%

{\footnotesize \textit{Note.} M~= mean number of splits; SD~= standard deviation of the number of splits; MAD~= mean absolute deviation from true tree size. Gray circles represent counts, dark gray horizontal lines represent true number of splits~3. Panel labels indicate whether variances of the random intercept and slope were fixed to 0 or freely estimated. Distances on $y$-axis are on the log scale. cl.cov.~= cluster-level covariances employed in parameter stability tests; ran.eff.~= estimation initialized with the random effects on the full sample. }
\label{fig:LMM_sizes}
\end{figure}

Figure~\ref{fig:LMM_sizes} depicts the number of splits implemented by each partitioning approach. The default fitting approach tends to overfit, irrespective of the random-effects specification, implementing $> 3$ splits in 89\% datasets, on average. The use of cluster-level covariances in the parameter stability tests successfully mitigated overfitting: LMM trees with clustered covariances with (middle panel) or without (right panel) random slopes showed an average number of splits closest to the true tree size. 

In terms of MAD, however, LMM trees initializing estimation with the random effects performed best, but only if the random slope was not estimated. Thus, initializing estimation with the random effects may only be useful if the random-effects specification is kept relatively simple. With more complex random-effects specifications, actual subgroup differences may be more likely captured by the random effects than by the tree structure. The second-lowest MAD was observed for LM trees with clustered covariances (left panel). Combined use of random-effects initialization and cluster-level covariances was not very effective, irrespective of the random-effects specification.

Distributions of the number of splits, separated according to the levels of the data-generating design are presented in Figure~\ref{fig:LMM_sizes_interact} and discussed in Appendix~\ref{sec:AppendixA}. The results show a pattern very similar to Figure~\ref{fig:LMM_sizes}; no substantial interactions between data-generating and model-specification parameters were observed. Main effects of the data-generating parameters were as expected: Strongest effects were for $\sigma_{b_1}$ and $N$, with higher values resulting in a higher number of splits. The number of noise variables and $\sigma^2_{b_1}$ had smaller effects, while the correlation between partitioning variables hardly affected the number of implemented splits.

Table~\ref{tab:first_splits} shows the variables selected for the first split, and indicates high accuracy for recovery of the first split for all LM(M) tree fitting approaches. Only very rarely is $u_1$ not selected for the first split, if a first split was implemented.

\begin{table}

\begin{threeparttable}
\caption{\label{tab:first_splits}Variables selected for the first split by each LM tree (top two rows) and LMM tree (bottom eight rows) estimation approach.}
\centering
\fontsize{11}{13}\selectfont
\begin{tabular}[t]{llccccc}
\toprule
Random effects & Fitting approach & $u_1$ & $u_2$ & $u_3$ & $u_4$--$u_{25}$ & No split\\
\midrule
$\hat{\sigma}_{b_0} = \hat{\sigma}_{b_1} = 0$ & default & 1.000 & 0.000 & 0.000 & 0.000 & 0.000\\
$\hat{\sigma}_{b_0} = \hat{\sigma}_{b_1} = 0$ & cl.cov. & 1.000 & 0.000 & 0.000 & 0.000 & 0.000\\
\midrule
$\hat{\sigma}_{b_0} > 0$, $\hat{\sigma}_{b_1} = 0$ & default & 1.000 & 0.000 & 0.000 & 0.000 & 0.000\\
$\hat{\sigma}_{b_0} > 0$, $\hat{\sigma}_{b_1} = 0$ & cl.cov. & 1.000 & 0.000 & 0.000 & 0.000 & 0.000\\
$\hat{\sigma}_{b_0} > 0$, $\hat{\sigma}_{b_1} = 0$ & ran.eff. & 0.996 & 0.002 & 0.002 & 0.001 & 0.000\\
$\hat{\sigma}_{b_0} > 0$, $\hat{\sigma}_{b_1} = 0$ & cl.cov. + ran.eff. & 1.000 & 0.000 & 0.000 & 0.000 & 0.000\\
\midrule
$\hat{\sigma}_{b_0} > 0$, $\hat{\sigma}_{b_1} > 0$ & default & 1.000 & 0.000 & 0.000 & 0.000 & 0.000\\
$\hat{\sigma}_{b_0} > 0$, $\hat{\sigma}_{b_1} > 0$ & cl.cov. & 1.000 & 0.000 & 0.000 & 0.000 & 0.000\\
$\hat{\sigma}_{b_0} > 0$, $\hat{\sigma}_{b_1} > 0$ & ran.eff. & 0.499 & 0.007 & 0.006 & 0.003 & 0.485\\
$\hat{\sigma}_{b_0} > 0$, $\hat{\sigma}_{b_1} > 0$ & cl.cov. + ran.eff. & 1.000 & 0.000 & 0.000 & 0.000 & 0.000\\
\bottomrule
\end{tabular}
\begin{tablenotes}
\small
\item [] \footnotesize \\ \textit{Note.} $u_1$ is the true first splitting variable and is binary; all other partitioning variables are continuous, with $u_2$ and $u_3$ being true splitting variables (nodes 2 and 3). $\hat{\sigma}_{b_0}$ and $\hat{\sigma}_{b_1}$ are the estimated standard deviations of the random intercept and slope, respectively.
\end{tablenotes}
\end{threeparttable}
\end{table}
\FloatBarrier

\section{Study~II: Comparison with Other Partitioning Methods}

Next, we compared the performance of LM(M) trees with that of SEM trees and LongCART. This allowed for evaluating the possible (dis)advantages of global versus local estimation of random-effects parameters, as well as the performance of the different splitting criteria employed by each method. The same data-generating design as in Simulation Study~I was employed. To reduce the number of comparisons, we only include performance of LM(M) trees fitted using clustered covariances, because they showed good performance in Simulation Study~I.

\subsection{Method}

We fitted a total of six SEM trees to every dataset. We used two different splitting criteria: 

\begin{itemize} 

\item The default ``naive'' splitting approach which employs likelihood ratio tests (LRTs) as the splitting criterion \citep{BranyOert13}. That is, for each candidate split, the log-likelihood of the SEM fitted to the observations in the current node is compared against the sum of the log-likelihoods of a two-group SEM, in which the two groups are defined by the candidate split. An LRT can thus be computed for each candidate split, which quantifies the improvement in fit that would result from implementing this split. In each step, the candidate split  yielding the highest LRT is selected for splitting, and splitting is continued as long as a candidate split yields a $p$-value of the LRT above a pre-specified $\alpha$ level (0.05, by default). 

\item The score-based splitting approach of \cite{ArnoyVoel21}. This approach uses the MOB algorithm described in the Introduction, where the parametric model fitted in step (a) is a SEM. While for GLMM trees, parameter stability tests are computed for the fixed-effects parameters only, score-based SEM trees compute parameter stability tests based on both fixed- and random-effects parameters. 

\end{itemize}

For each splitting criterion, three different random-effects specifications were employed:
\begin{itemize}
  \item None: $\hat{\sigma}_{b_0} = \hat{\sigma}_{b_1} = 0$. Random effects were not estimated and their variances were thus fixed to 0.
  
  \item Intercepts: $\hat{\sigma}_{b_0} > 0; \hat{\sigma}_{b_1} = 0$. In every node, the variance of the random intercept was freely estimated; the variance of the random slope was fixed to 0. 
  
  \item Intercepts and slopes: $\hat{\sigma}_{b_0} > 0; \hat{\sigma}_{b_1} > 0$. In every node, the variances of the random intercept and slope, as well as their correlation, were freely estimated. 
\end{itemize}
To specify the node-specific models for SEM trees, we employed an LGCM specification with the response at each timepoint regressed on a latent intercept and slope. Intercept loadings were fixed to 1; slope loadings were fixed to ${0, 1, 2, 3, 4}$, respectively. Errors were assumed uncorrelated between timepoints and an error variance was freely estimated for each timepoint. We used package \textbf{lavaan} \citep[version 0.6-11; ][]{Ross12} to fit the SEMs and we used package \textbf{semtree} \citep[version 0.9.17; ][]{BranyOert13} to fit the SEM trees.    

We fitted a single LongCART tree to each dataset. The LongCART function estimates node-specific models comprising a random intercept term; this default cannot be changed. A fixed-effects model was specified with the response regressed on time and a subject-specific random intercept. We used package \textbf{LongCART} \citep[version 3.1][]{Kund21} to fit LongCART trees.

\subsection{Results}

\subsubsection{Tree size}

\begin{figure}[tb]
\caption{Tree size distributions for LM(M) trees with clustered covariances, SEM trees and LongCART.}
\begin{subfigure}{1.25\textwidth}
\includegraphics{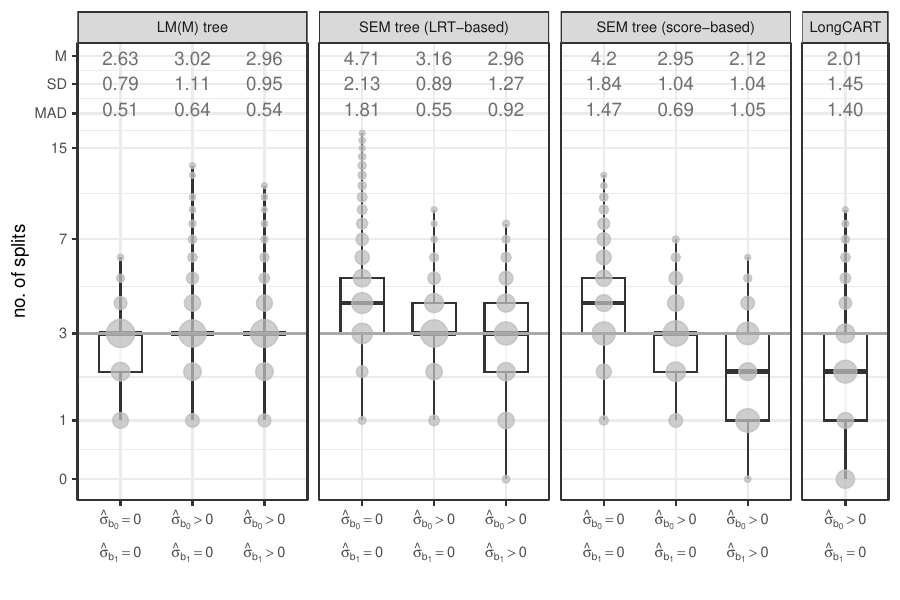}
\end{subfigure}%
\vspace{-0.5cm}%
{\footnotesize \textit{Note.} M~= mean number of splits; SD~= standard deviation of the number of splits; MAD~= mean absolute deviation from true tree size. Gray circles represent counts, dark gray horizontal lines represent true number of splits~3. Distances on the $y$-axis are on the log scale; $x$-axis labels indicate whether variances of the random intercept and slope were fixed to 0 or freely estimated.}
\label{fig:tree_sizes}
\end{figure}

Figure~\ref{fig:tree_sizes} depicts tree size distributions for the different algorithms. SEM trees (middle two panels) performed well, exhibiting overfitting only when no random effects were specified. LRT-based SEM trees tended to implement more splits than score-based SEM trees. Both LRT- and score-based SEM trees yielded lower tree size with increasing complexity of the random-effect specification. With the correct random-effects specification (i.e., both random intercept and slope were estimated), LRT-based SEM trees showed average tree size very close to the true tree size, while under-specification (i.e., specification of only a random intercept) increased tree size by 0.2 splits on average. Score-based SEM trees perform best when only random intercepts are estimated, implementing too few splits when random slopes were (correctly) specified. LongCART trees seem to suffer from a lack of power, implementing only two splits on average. Overall, LM(M) trees with clustered covariances showed best performance, but are very closely followed by LRT-based SEM trees, which seem to be affected more strongly by mis-specification of the random effects.

Distributions of the number of splits, separated according to the levels of the data-generating parameters are depicted and discussed in Appendix~\ref{sec:AppendixA} and Figure~\ref{fig:tree_sizes_interact}. They are omitted here, as they show a pattern very similar to Figure~\ref{fig:tree_sizes}. Of the four data-generating parameters, $N$ and $\sigma_{b_0}^2$ showed the strongest effects, with higher values resulting in a higher number of splits, as expected. The number of splits implemented by SEM trees was most strongly affected by the data-generating parameters when the random effects were mis-specified (i.e., random intercepts and/or slope fixed to 0).

\FloatBarrier

\subsubsection{Split selection}

\begin{table}

\begin{threeparttable}
\caption{\label{tab:first_splits2}Variables selected for the first splits by each of the partitioning approaches.}
\centering
\fontsize{11}{13}\selectfont
\begin{tabular}[t]{llccccc}
\toprule
Algorithm & Random effects & $u_1$ & $u_2$ & $u_3$ & $u_4$--$u_{25}$ & No split\\
\midrule
LM(M) tree & $\hat{\sigma}_{b_0} = \hat{\sigma}_{b_1} = 0$ & 1.000 & 0.000 & 0.000 & 0.000 & 0.000\\
(cl.cov.) & $\hat{\sigma}_{b_0} > 0$, $\hat{\sigma}_{b_1} = 0$ & 1.000 & 0.000 & 0.000 & 0.000 & 0.000\\
 & $\hat{\sigma}_{b_0} > 0$, $\hat{\sigma}_{b_1} > 0$ & 1.000 & 0.000 & 0.000 & 0.000 & 0.000\\
SEM tree & $\hat{\sigma}_{b_0} = \hat{\sigma}_{b_1} = 0$ & 1.000 & 0.000 & 0.000 & 0.000 & 0.000\\
(LRT-based) & $\hat{\sigma}_{b_0} > 0$, $\hat{\sigma}_{b_1} = 0$ & 0.998 & 0.000 & 0.002 & 0.000 & 0.000\\
 & $\hat{\sigma}_{b_0} > 0$, $\hat{\sigma}_{b_1} > 0$ & 0.992 & 0.001 & 0.001 & 0.002 & 0.004\\
SEM tree & $\hat{\sigma}_{b_0} = \hat{\sigma}_{b_1} = 0$ & 0.761 & 0.176 & 0.052 & 0.012 & 0.000\\
(score-based) & $\hat{\sigma}_{b_0} > 0$, $\hat{\sigma}_{b_1} = 0$ & 0.879 & 0.077 & 0.044 & 0.000 & 0.000\\
 & $\hat{\sigma}_{b_0} > 0$, $\hat{\sigma}_{b_1} > 0$ & 0.999 & 0.000 & 0.000 & 0.000 & 0.001\\
LongCART & $\hat{\sigma}_{b_0} > 0$, $\hat{\sigma}_{b_1} = 0$ & 0.000 & 0.419 & 0.298 & 0.092 & 0.192\\
\bottomrule
\end{tabular}
\begin{tablenotes}
\small
\item [] \footnotesize \\ \textit{Note.} $u_1$ is the true first splitting variable and is a binary factor; all other partitioning variables are continuous, with $u_2$ and $u_3$ being true splitting variables (nodes 2 and 3). The first column indicates wheter the random intercept and/or slope were estimated or not.
\end{tablenotes}
\end{threeparttable}
\end{table}

Table~\ref{tab:first_splits2} presents variable selection frequencies for the first split in the fitted trees. For SEM trees, the LRT criterion yields almost perfect accuracy for the first split. The score-based criterion for SEM trees provides near-perfect accuracy when random effects were correctly specified. When random effects were mis-specified, score-based SEM trees selected the wrong variable for the first split in about 12\% of datasets. Closer inspection of stability tests for individual models and parameters suggested that the the score-based tests for SEMs are more sensitive to instability in the fixed slope than in the fixed intercept, explaining why $u_2$ or $u_3$ were often selected for the first split.

LongCART trees exhibit low accuracy for recovering the first split, selecting the wrong variable in all datasets where at least one split was implemented. LongCART showed a strong tendency to select $u_2$ or $u_3$ for the first split. Closer inspection of the fitted LongCART trees revealed that in 99\% percent of datasets in which no splits were implemented, $u_1$ was the strongest splitting candidate, but the parameter stability tests did not reach significance. This suggests the tests proposed by \cite{KundyHare19} are less sensitive to instability of the intercept (compared to instability of the slope), or less sensitive to instability with respect to categorical covariates (compared to instability with respect to continuous covariates).

\subsubsection{Computation time}

\begin{figure}[!ht]%
\caption{Computation time distributions for the different partitioning methods.}
\begin{subfigure}{.8\textwidth}%
\includegraphics{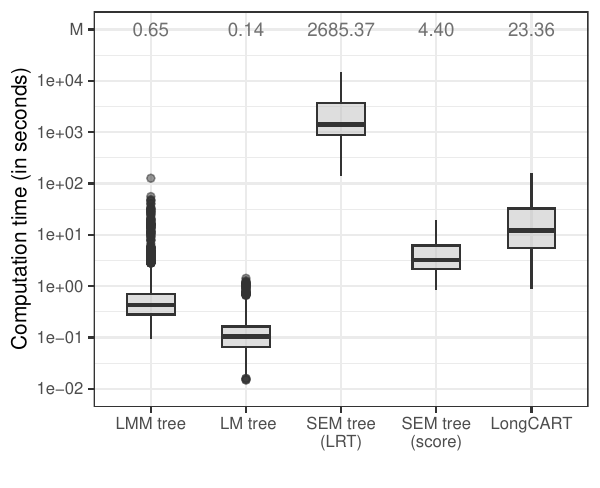}
\end{subfigure}%
\\ {\footnotesize \textit{Note.} $y$-axis is on the log scale; M = mean computation time in seconds. }
\label{fig:comp_times}
\end{figure}%

Figure~\ref{fig:comp_times} presents computation time distributions for the partitioning algorithms. A clear computational advantage is observed for LM trees. LMM trees require longer computation times because of the estimation of random effects. Yet, this increase is minor compared to the computation times required by LongCART and SEM trees: LRT-based SEM trees required striking computation times, while score-based SEM trees were computationally more efficient, as expected.

\FloatBarrier

\section{Study~III: Partitioning Academic Trajectories}

\subsection{Method} 

\subsubsection{Dataset}

We analyzed trajectories on children's reading, math and science abilities from the Early Childhood Longitudinal Study-Kindergarten class of 1998--1999 \citep[ECLS-K; ][]{NCES10}. Data were collected from 21,304 children from 1,018 schools across the USA. Assessments took place from kindergarten in 1998 through 8th grade in 2007, here we focus on assessments from kindergarten, 1st, 3rd, 5th and 8th grade.

Response variables are reading, math, and science abilities, which were assessed using multi-item cognitive tests. Latent ability estimates were computed with a mean of zero and variance of one. Reading and math abilities were assessed in all five rounds of data collection, science knowledge was assessed in 3rd, 5th and 8th grade. We analyzed data from children who completed all assessments yielding $N = 6,277$ for reading; $N = 6,512$ for math; $N = 6,625$ for science.

Time was measured as the number of months since the baseline assessment. In order to obtain approximately linear trajectories, we chose the timing metric based on visual inspection of the observed data: $\mathrm{months}^{1/2}$ was used as the timing metric for reading and math trajectories, and $\mathrm{months}^{2/3}$ for science trajectories.

We used 11 time-invariant covariates as potential partitioning variables, all assessed at baseline: Gender (51.1\% male); Age in months (range 53 to 96; M = 6.14 years); Race (8 categories); First time in kindergarten (yes/no); Socio-economic status (range $-$5 to 3); Fine motor skills (e.g., drawing figures; range 0 to 9); Gross motor skills (e.g., ability to hop, skip and jump; range 0 to 8); Interpersonal skills (range 1 to 4); Self-control (range 1 to 4); Internalizing problem behavior (range 1 to 4); Externalizing problem behavior (range 1 to 4).

\subsubsection{Fitting approaches}

We applied five LM(M) trees to the data, focusing on the original GLMM tree approach and approaches that performed well in the simulations:
\begin{itemize}
  \item An LM tree ($\hat{\sigma}_{b_0} = \hat{\sigma}_{b_1} = 0$) using clustered covariances in the parameter-stability tests.
  \item An LMM tree using observation-level covariances, with a random intercept freely estimated ($\hat{\sigma}_{b_0} > 0$, $\hat{\sigma}_{b_1} = 0$) and initialization assuming zero random effects.
  \item An LMM tree using clustered covariances, with a random intercept freely estimated ($\hat{\sigma}_{b_0} > 0$, $\hat{\sigma}_{b_1} = 0$) and initialization assuming zero random effects.
  \item An LMM tree using clustered covariances, with both random intercept and slope freely estimated ($\hat{\sigma}_{b_0} > 0$, $\hat{\sigma}_{b_1} > 0$) and initialization assuming zero random effects . 
  \item An LMM tree using observation-level covariances, with a random intercept freely estimated ($\hat{\sigma}_{b_0} > 0$, $\hat{\sigma}_{b_1} = 0$) and random-effect initialization with the full-sample estimate. 
\end{itemize}
Although LRT- and score-based SEM trees performed very well in the simulations, they could not be used in this study because growth-curve SEMs do not allow for incorporating continuous time. 

%
%

%
%

\subsubsection{Evaluation of performance}

The ECLS-K datasets have exceptionally large sample sizes, so we employed random sampling to obtain training samples of $N=250$ children, likely more representative of real-world studies in psychology. We performed 100 repetitions for each response variable (math, reading, or science). We evaluated predictive accuracy by computing the mean squared difference between predicted and observed response variable values (MSE) for all children not included in the training sample in the current repetition. This separation of train and test observations does not allow for using the random effects in computing predictions; the cross-validated MSEs only quantify accuracy of the fixed-effects parameters. Tree size was measured by counting the number of splits in each tree.

\FloatBarrier
\subsection{Results}

\begin{figure}[!bt]
\caption{Mean squared errors for trees fitted to math, reading and science ability trajectories.}
\begin{subfigure}{\textwidth}
\includegraphics{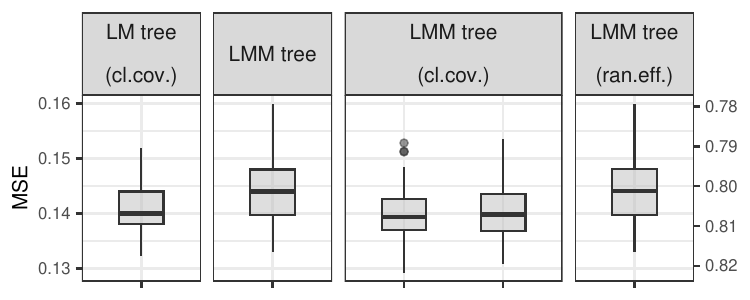}
\end{subfigure}
\begin{subfigure}{\textwidth}
\includegraphics{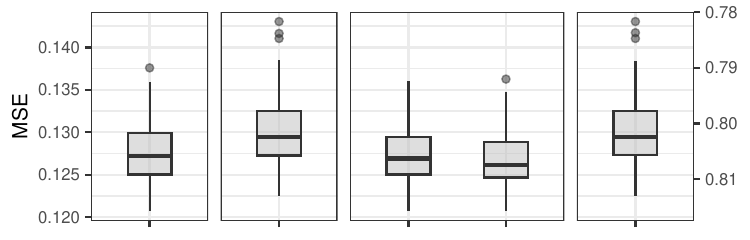}
\end{subfigure}
\begin{subfigure}{\textwidth}
\includegraphics{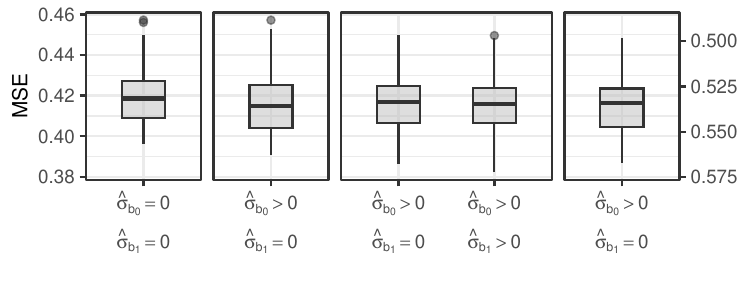}
\end{subfigure}%
\vspace{-0.5cm}%
{\footnotesize \\ \textit{Note. } Top, middle and bottom panels depict math, reading and science ability trajectories, respectively. The secondary $y$-axis on the right quantifies the proportion of variance explained, computed as $1 - \frac{\text{mean(MSE)}}{\text{var}(y)}$.}
\label{fig:application_MSEs}
\end{figure}

Figure~\ref{fig:application_MSEs} and Table~\ref{tab:application_MSEs} present MSE distributions. Differences in predictive performance are small, all $R^2$ differences are smaller than 0.01. Bonferroni-adjusted pairwise $t$-tests indicated no difference in performance between default LMM trees and LMM trees with random-effects initialization for any of the three outcomes. In contrast, the three LM(M) trees using clustered covariances performed significantly better for the math and reading outcomes. For the science outcomes, no significant differences were observed. Thus, cluster-level covariances seem to provide the most robust improvement of performance. 

\begin{figure}[!bt]%
\caption{Sizes of trees fitted to math, reading, and science ability trajectories.}
\begin{subfigure}{\textwidth}%
\includegraphics{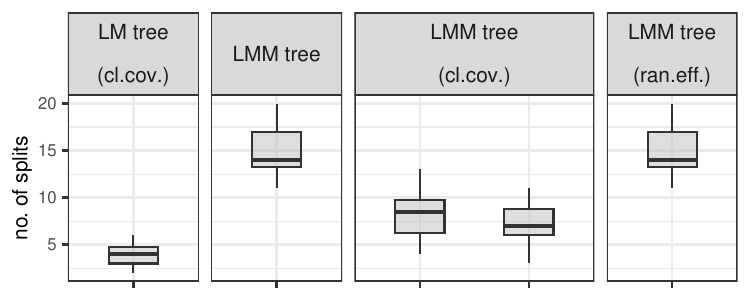}
\end{subfigure}
\begin{subfigure}{\textwidth}
\includegraphics{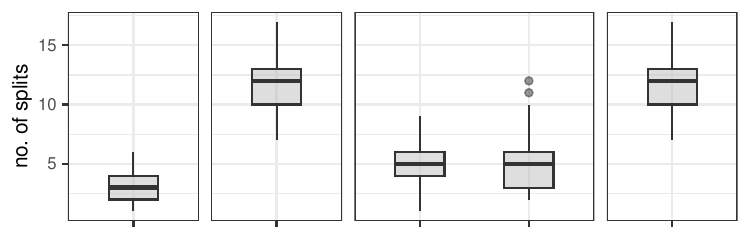}
\end{subfigure}
\begin{subfigure}{\textwidth}
\includegraphics{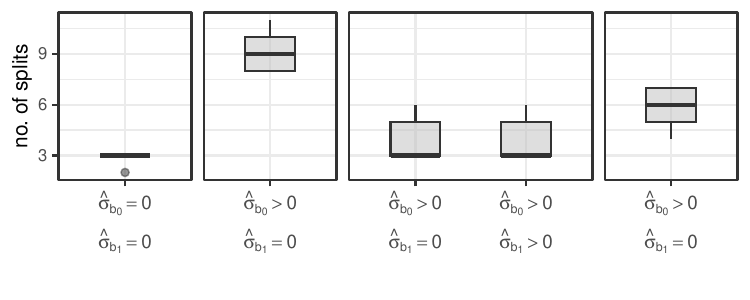}
\end{subfigure}%
{\\\footnotesize \textit{Note.} Top, middle and bottom panels depict math, reading and science ability trajectories, respectively.}
\label{fig:application_sizes}
\end{figure}

Figure~\ref{fig:application_sizes} presents tree size distributions. Similar to the simulation study's results, default LMM trees implement the largest number of splits. Cluster-level covariances provide a robust reduction in the number of splits. Given their non-distinguishable predictive performance, LM trees with clustered covariances may be preferred if a obtaining a sparse result is critical. LMM trees with clustered covariances may provide less sparsity but better predictive accuracy, irrespective of whether random intercepts and/or slopes were estimated. In contrast to the simulation study results, random-effects initialization here yields tree sizes similar to the default fitting approach.

\begin{table}

\begin{threeparttable}
\caption{\label{tab:application_MSEs}Cross-validated mean squared errors for each of the response variables.}
\centering
\fontsize{11}{13}\selectfont
\begin{tabular}[t]{lccccccccc}
\toprule
\multicolumn{1}{c}{ } & \multicolumn{3}{c}{Math} & \multicolumn{3}{c}{Reading} & \multicolumn{3}{c}{Science} \\
\cmidrule(l{3pt}r{3pt}){2-4} \cmidrule(l{3pt}r{3pt}){5-7} \cmidrule(l{3pt}r{3pt}){8-10}
  & M & SD & $R^2$ & M & SD & $R^2$ & M & SD & $R^2$\\
\midrule
LM tree$^c$ & 0.1407 & 0.004 & 0.806 & 0.1274 & 0.003 & 0.806 & 0.4190 & 0.013 & 0.531\\
LMM tree$^i$ & 0.1441 & 0.006 & 0.801 & 0.1302 & 0.004 & 0.801 & 0.4154 & 0.014 & 0.535\\
LMM tree$^{i,c}$ & \textbf{0.1398} & \textbf{0.004} & \textbf{0.807} & 0.1273 & 0.003 & 0.806 & 0.4163 & 0.012 & 0.534\\
LMM tree$^{i,s,c}$ & 0.1403 & 0.005 & 0.806 & \textbf{0.1269} & \textbf{0.003} & \textbf{0.806} & 0.4156 & 0.013 & 0.535\\
LMM tree$^{i,r}$ & 0.1441 & 0.006 & 0.801 & 0.1302 & 0.004 & 0.801 & \textbf{0.4151} & \textbf{0.012} & \textbf{0.536}\\
\bottomrule
\end{tabular}
\begin{tablenotes}
\small
\item [] \footnotesize \\ \textit{Note.} Means and standard deviations computed over 100 cross-validation repetitions. Boldfaced values indicate the best-performing method for each outcome. $R^2$ was computed as $\frac{\text{mean(MSE)}}{\text{var}(y)}$. $^c$~cluster-level covariances; $^r$~estimation initialized with random effects; $^i$~random-intercept variance freely estimated; $^s$~random-slope variance freely estimated.
\end{tablenotes}
\end{threeparttable}
\end{table}

\FloatBarrier

\section{Discussion}

The simulations showed that the proposed extensions of GLMM trees are effective for partitioning LGCMs. Use of clustered covariances appears most effective and their good performance appears largely unaffected by (mis-)specification of the random effects; it may therefore be the safest choice in practice. Initializing estimation with the random effects was also effective, but only when the random-effects specification is kept simple (i.e., no estimation of random slopes). Combining cluster-level covariances and random-effects initialization worsened performance and is thus not recommended. 

Strong performance of clustered covariances was also observed in partitioning real-world academic trajectories. They provided substantially smaller trees for all outcomes and better or equal predictive accuracy. In comparison to cluster-level covariances, random-effects initialization resulted in larger trees for all outcomes and worse performance for the reading and math outcomes.

The simulations showed comparable performance of LM(M) and SEM trees in partitioning LGCMs. SEM trees may however be more sensitive to mis-specification of the random effects, with under-specification resulting in too many splits. In line with results of \cite{ArnoyVoel21}, we found score-based SEM trees to have somewhat lower power than LRT-based SEM trees, but at a much lower computational cost. LongCART trees often selected the wrong partitioning variable for the first split, and were outperformed by LM(M) and SEM trees. The LongCART parameter stability tests \citep{KundyHare19} may be underpowered for detecting instability of the fixed intercept, or for detecting instability with respect to categorical covariates. 

The simulations clearly illustrated the lower computational burden of GLMM trees. This is in large part due to their local-global estimation approach, where fixed-effects parameters are estimated locally within a node and random-effects parameters are estimated globally, using all observations. In contrast, SEM trees and LongCART fit the full mixed-effects model in each node, which substantially increases computational load. The local-global estimation approach also reduces model complexity, because a lower number of random-effects parameters need to be estimated.

Yet, a possible downside of the local-global estimation approach is that it does not allow for recovering subgroups with differences in random-effects parameters. When there is a specific interest in partitioning the random effects, score-based SEM trees may be preferred. Alternatively, researchers may want to use the parameter stability tests for mixed-effects models developed by \cite{WangyMerk18} \citep[see also ][]{WangyMerk21,WangyGrav22}. This will be useful, for example, when the number of or distances between timepoints differ between respondents and SEM-based growth curve models cannot be applied \citep{NeisyMatt18}.

The current evaluations were limited to Gaussian responses and LGCMs. Future studies should assess performance of GLMM trees in partitioning longitudinal data with, for example, binomial or count responses. We expect that the strong performance of cluster-level covariances generalizes to other settings where covariates are measured at higher levels, either in longitudinal and/or otherwise nested data structures, but whether our conclusions generalize to settings beyond LGCMs remains to be evaluated. Finally, we used the outer-product-of-gradients (OPG) estimator for computing (clustered) covariances. Though computationally more burdensome, future work could assess potential benefits of using the full sandwich estimator.

\bibliography{bib}

\newpage
\appendix

\section{Appendix A: Effects of data-generating parameters on tree size}
\label{sec:AppendixA}

\begin{figure}[!ht]
\caption{Effects of data-generating parameters on tree size for LM(M) trees.}
\begin{subfigure}{1.25\textwidth}
\includegraphics{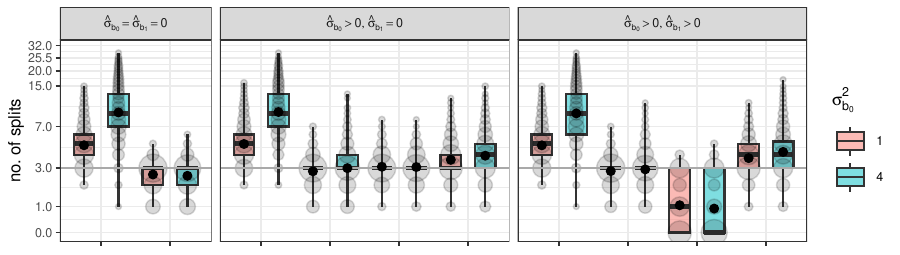}
\end{subfigure}
\begin{subfigure}{1.25\textwidth}
\includegraphics{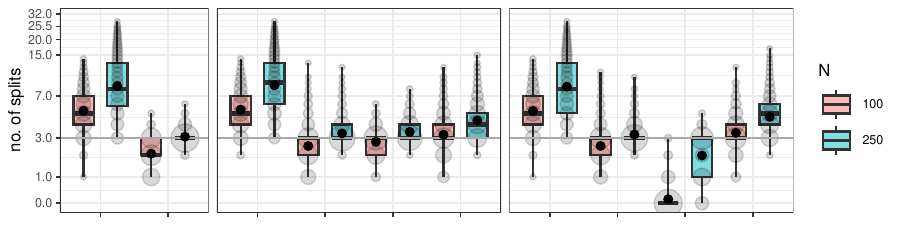}
\end{subfigure}
\begin{subfigure}{1.25\textwidth}
\includegraphics{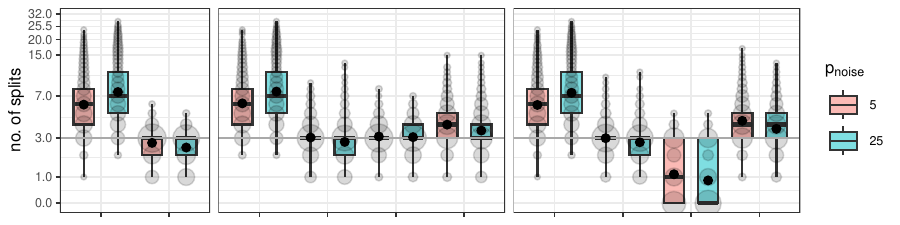}
\end{subfigure}
\begin{subfigure}{1.25\textwidth}
\includegraphics{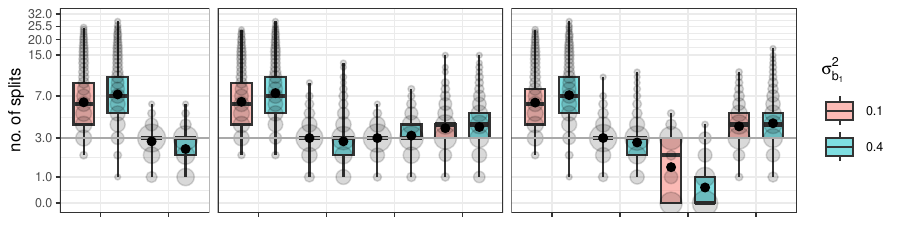}
\end{subfigure}
\begin{subfigure}{1.25\textwidth}
\includegraphics{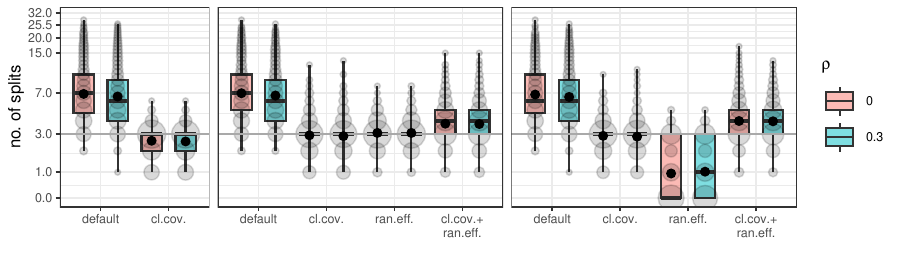}
\end{subfigure}%
\vspace{-.6cm}%
{\singlespacing \footnotesize \textit{Note.} Black dots represent means, gray circles represent counts, dark gray horizontal lines represent true number of splits~3. Distances on $y$-axis are on log scale. $\sigma_{b_0}^2$ = variance of random intercept; $\sigma_{b_1}^2$ = variance of random slope; $N$ = sample size at level 2; $p_{noise}$ = number of noise variables; $\rho$ = correlation between partitioning variables.}
\label{fig:LMM_sizes_interact}
\end{figure}

For LM(M) trees (Figure~\ref{fig:LMM_sizes_interact}), use of cluster-level covariances provided the most robust improvement in split recovery. For the data-generating parameters, the effects on tree size were strongest for $\sigma_{b_0}$, followed by $N$, $p$, $\sigma_{b_1}$ and $\rho$. Higher values of $\sigma_{b_0}$, $N$, $p$ and $\sigma_{b_1}$ tend to yield higher numbers of splits, while the effect of $\rho$ is minimal. 

\begin{figure}[!ht]
\caption{Effect of data-generating parameters on tree size for LM(M), SEM and LongCART trees.}
\begin{subfigure}{1.25\textwidth}
\includegraphics{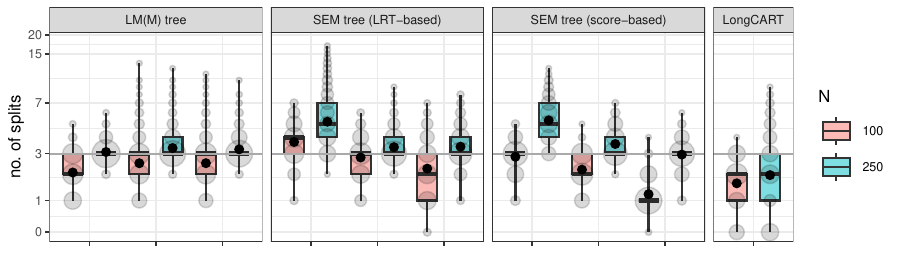}
\end{subfigure}
\begin{subfigure}{1.25\textwidth}
\includegraphics{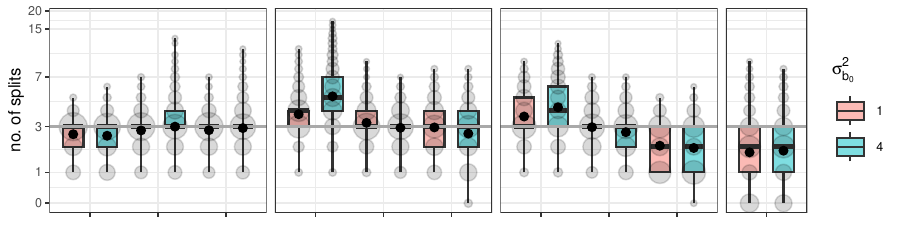}
\end{subfigure}
\begin{subfigure}{1.25\textwidth}
\includegraphics{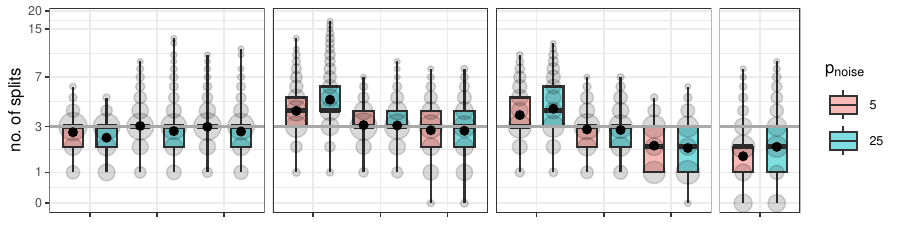}
\end{subfigure}
\begin{subfigure}{1.25\textwidth}
\includegraphics{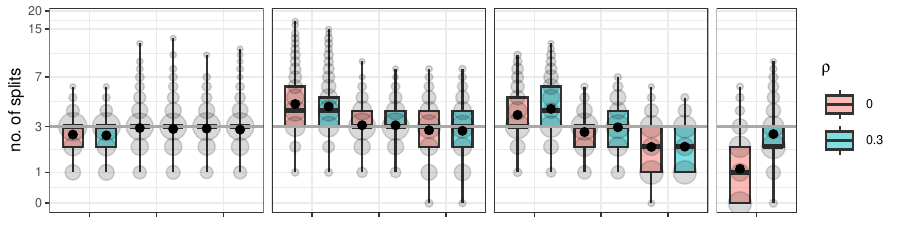}
\end{subfigure}
\begin{subfigure}{1.25\textwidth}
\includegraphics{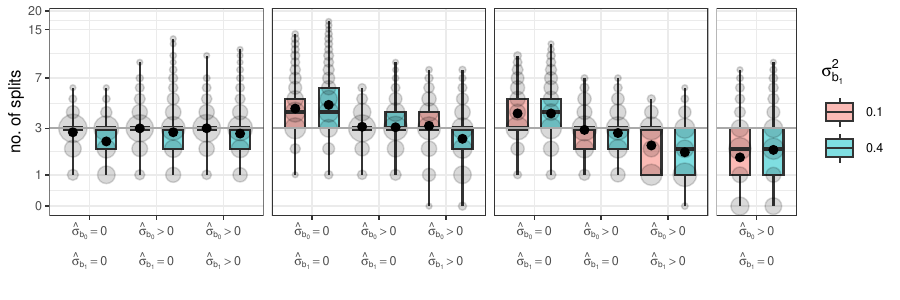}
\end{subfigure}%
\vspace{-.6cm}%
{ \singlespacing \footnotesize \textit{Note.} Black dots represent means, gray circles represent counts, dark gray horizontal lines represent true number of splits~3. Distances on $y$-axis are on log scale. $N$ = sample size at level 2; $\sigma_{b_0}^2$ = variance of random intercept; $\sigma_{b_1}^2$ = variance of random slope; $p_{noise}$ = number of noise variables; $\rho$ = correlation between partitioning variables.}
\label{fig:tree_sizes_interact}
\end{figure}

For SEM trees and LongCART (Figure~\ref{fig:tree_sizes_interact}), strongest effects were observed for $N$, followed by $\sigma_{b_0}$, $p_{noise}$, $\rho$ and $\sigma_{b_1}$. Results for $N$ were as expected for all methods: More splits are implemented with higher sample size. Both LRT- and score-based SEM trees seem only affected by levels of $\sigma_{b_0}$, $p_{noise}$ and $\rho$ under misspecification of the random effects. Especially when both random intercept and slope variances are fixed to 0, higher levels of $\sigma_{b_0}$ and $p_{noise}$ yield more splits with both SEM tree approached. LRT-based SEM trees seem unaffected by levels of $\rho$, while score-based SEM trees seemed to implement a larger number of splits with increasing values of $\rho$. This pattern seemed reversed for increased magnitude of $\sigma_{b_1}$, which yields a lower number of splits for both SEM tree approaches, but only when the random effects were correctly specified. LongCART implemented more splits with higher levels of $\rho$,  $\sigma_{b_1}$, $N$, $p_{noise}$ and $\sigma_{b_0}$.

\FloatBarrier

\end{document}